\documentclass[aps,pra,twocolumn,groupedaddress,showpacs]{revtex4}
\usepackage{graphicx}
\graphicspath{{figure file fold path}}
\usepackage[dvipsnames,usenames]{color}
\usepackage{color}
\usepackage{amsmath}
\usepackage{amssymb}

\emergencystretch=\maxdimen
\begin{document}

\title{Photon statistics and dynamics of nanolasers subject to intensity feedback}

\author{T.~Wang$^{1}$, Z.L.~Deng$^{1}$, J.C.~Sun$^{1}$, X.H.~Wang$^{1}$, G.P.~Puccioni$^{2}$, G.F.~Wang$^{1}$, and G.L.~Lippi$^{3}$}
\affiliation{$^1$School of Electronics and Information, Hangzhou Dianzi University, Hangzhou 310018, China}
\affiliation{$^3$Istituto dei Sistemi Complessi, CNR, Via Madonna del Piano 10, I-50019 Sesto Fiorentino, Italy}
\affiliation{$^2$Universit\'e C\^ote d'Azur, Institut de Physique de Nice (INPHYNI), CNRS UMR 7010, Nice, France}




\begin{abstract}

Using a fully stochastic numerical scheme, we theoretically investigate the behaviour of a nanolaser in the lasing transition region under the influence of intensity feedback. Studying the input-output curves as well as the second order correlations for different feedback fractions, we obtain an insight on the role played by the fraction of photons reinjected into the cavity. Our results reveal that the transition shrinks and moves to lower pump values     with the feedback strength, and irregular pulses can be generated when feedback is sufficiently large. The interpretation of the observation is strengthened through the comparison with the temporal traces of the emitted photons and with the radiofrequency power spectra. These results give insight into the physics of nanolasers as well as validating the use of the second order autocorrelation as a sufficient tool for the interpretation of the dynamics. This confirmation offers a solid basis for the reliance on autocorrelations in experiments studying the effects of feedback in nanodevices.

\end{abstract}

\maketitle

\section{Introduction}
The continuous downscaling of laser cavity size has opened the door to the realization of nanoscale laser sources that are based on the altered spontaneous emission of the gain material~\cite{Wiersig2009, Hostein2010}. Such compact light sources can be densely integrated on-chip with potential applications in optical communication~\cite{Smit2012,Crosnier2017}, high-speed optical computing~\cite{Hill2014}, nonlinear optical microscopy~\cite{Nakayama2007,Deeb2018}, and even sensing~\cite{Shambat2013, Nechepurenko2018}. In particular, the ultra-fast responsivity and low power consumption of all-optical or hybrid computing can save hundreds of terawatt-hours per year: a significant portion of the global energy use~\cite{Pan2018}. Therefore, efforts in laser miniaturization have concentrated on the improvement and investigation of nanolasers with a large fraction of spontaneous emission into the lasing mode ($\beta$). High-$\beta$ lasers can theoretically achieve ultra-low threshold since most of the spontaneous emission is coupled into the lasing mode. In the limit $\beta = 1$ (the so-called \textit{thresholdless} regime) the laser output linearly increases with input power. Pursuing this ideal device - the one with the lowest power consumption and widest modulation bandwidth~\cite{Ding2015} - requires, however, finding the answer to some outstanding problems and the generalization of concepts which have been already investigated in macroscopic devices. One crucial point concerns the identification of the pump region for coherent emission and the possible exploitation of the photon output in the partially coherent regime~\cite{Wang2019}. Indeed, the proper exploitation of the ultra-low threshold of high-$\beta$ lasers can be most successful if the fully coherent emission regime does not need to be entirely achieved, since the latter requires pump rates which are comparatively larger than those required in macroscopic laser. For concreteness, if for $\beta = 1$ the amount of energy necessary for the achievement of coherence is $\varepsilon_c = 100\varepsilon_{th}$ (a realistic energy value), then the needed pump rate is the same as the one which corresponds to threshold for a $\beta = 10^{-2}$ device. Thus, operating a high-$\beta$ laser in the transition region, where the generation of trains of optical pulses has been demonstrated in mesoscale devices~\cite{Wang2019}, promises advantages whose properties need to be explored. In this contribution we therefore concentrate on the low coherence emission, below full-coherence operation~\cite{Lohof2018,Vyshnevyy2018,Takemura2019}.

Optical feedback, whether due to parasitic reflections or to built-in elements, is an additional ingredient whose influence on the performance of high-$\beta$ lasers needs to be studied in better detail. In macroscopic, low-$\beta$ lasers feedback has been the subject of extensive studies for over thirty years. In this case, it is well known that macroscopic semiconductor lasers display high sensitivity to external perturbations, and can achieve ``coherence collapse'' when the feedback is sufficiently strong~\cite{Mork1988, Cohen1989}. The interest in such a configuration, often related to numerous applications, arises from the rich phenomenology observed, ranging from increased noise, mode hopping, linewidth narrowing and broadening, and transition to developed chaos (coherence collapse)~\cite{Torcini2006}. Recent work in small-scale devices (``high-$\beta$'' lasers) subject to optical feedback reveals interesting physics, including chaos~\cite{Albert2011}, mode-switching\cite{Holzinger2018}, linewidth enhancement~\cite{Holzinger2018a} and various nonlinear dynamical phenomena~\cite{Hopfmann2013, Wang2019a}. However, a complete understanding of the physical mechanisms as the basis of this behaviour at the nanoscale still needs further investigation, given the simultaneous role played by deterministic rules -- which determine the dynamics -- and stochastic effects -- originating from the spontaneous processes -- in the behaviour of nanodevices~\cite{Wang2018ieee}.

In this paper, we numerically investigate the influence of external optical feedback onto the dynamics of nanolasers in the pump range corresponding to the threshold region. The numerical simulation of the nanolaser behaviour is conducted with the help of a fully stochastic, recursive approach~\cite{Puccioni2015} based on the semiclassical theory of stimulated emission~\cite{Einstein1917}. The advantage of the stochastic simulation (cf. also~\cite{Choudhury2009,Choudhury2010,Vallet2019}) is the automatic inclusion of two intrinsic noise sources:  the discreteness of the changes in photon and carrier number, shown to introduce dynamical fluctuations which cannot be captured by a differential description~\cite{Lebreton2013}, and the intrinsic stochasticity of all physical processes -- pumping, spontaneous emission, stimulated emission and transmission through the cavity mirrors.  The latter represent the physical origin of noise sources of non-gaussian nature, which can be approximated by gaussian distributions (as in the Langevin approach used to add noise to the Rate Equations~\cite{Coldren1995}) only whenever the photon number contained in an integration timestep is sufficiently large~\cite{Lippi2018,Lippi2019}.  This condition is fulfilled only when modeling a macroscopic laser ($\beta \lessapprox 10^{-5}$) above threshold.  Since in the present work we investigate smaller lasers ($\beta = 0.1$) in the pump region across threshold, the only adequate modeling currently available is a stochastic one, in spite of its shortcomings (absence of phase information).  As long as the feedback length is longer than the coherence length -- a condition easily met in high-$\beta$ devices -- the computations are correct. 

Our investigation extends experimental work, supported by numerics based on the same modelling scheme, carried out in a mesoscale laser but only for very weak feedback~\cite{Wang2019a}. Here, we focus our investigation on a nanolaser and numerically study the influence of different feedback levels. Anticipating on what follows, we find that irregular pulsing dynamics can be obtained for sufficiently large feedback. In addition, we find that the second order autocorrelation functions -- the powerful indicator currently available to characterize nanolaser dynamics -- are capable of capturing in an efficient way the characteristics of the pulsing dynamics, without, however, distinguishing between regular and irregular sequences.  

\section{Principle of optical feedback}
\begin{figure}[ht]
\includegraphics[width=0.90\linewidth,clip=true]{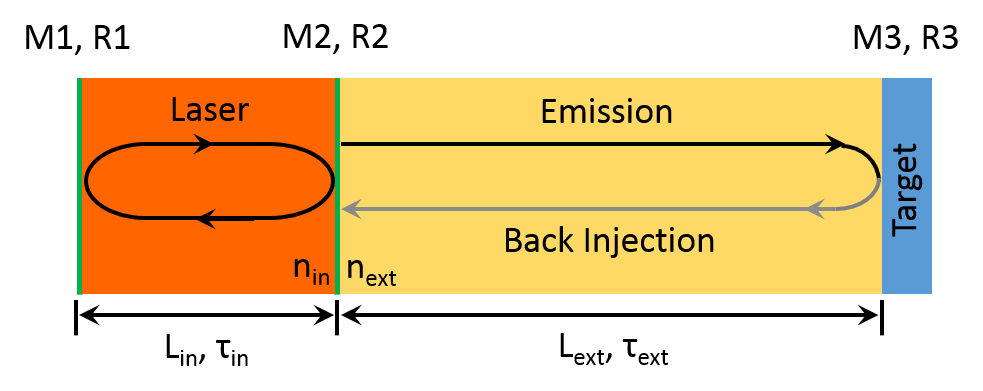}
\caption{Schematics of the laser subject to optical reinjection. The laser cavity is assumed to have a perfectly reflecting mirror (left) and an output coupler ($M_2$ with reflection coefficient for the photons $R_2$). The laser light propagates forward until hitting the target which reflects back a portion of the impinging photons. Only the prescribed fraction of photons can be coupled back into the cavity.}
\label{Archtec}
\end{figure}

The optical feedback scheme, illustrated in Fig.~\ref{Archtec}, is the typical one where a portion of the photons emitted by the semitransparent mirror $M_2$ is reflected back by the target ($M_3$ with reflectivity $R_3$), and thus propagates twice through the external cavity with length $L_{ext}$ (with corresponding delay time $\tau_{ext}$). In the single-mode models used for macroscopic lasers the fraction of the coherent optical field which reenters the cavity interferes with the intracavity one, possibly destabilizing the operation frequency (and coherence properties)~\cite{Lenstra1984,Cohen1989,Torcini2006} in addition to inducing variations in the output power with regular or irregular temporal dynamics~\cite{Kitaoka1996}.  This kind of dynamics rests on the coherent nature of the emission which, in macrolasers, takes place starting from threshold. This is not the case for small scale devices, where the emitted photons possess only a limited degree of coherence over a large range of pump values~\cite{Strauf2006,Wiersig2009,Hostein2010,Kreinberg2017,Ota2017,Kreinberg2019}. In the transition region, on which we are focussing, the optical reinjection is therefore (mostly) incoherent and can be treated on the basis of the uncoupled photon fraction~\cite{Wang2019a}. The inherent peculiarities of this regime rest on the strongly stochastic nature of the reinjection, which is properly captured by a fully stochastic simulation~\cite{Choudhury2009,Choudhury2010,Chusseau2014,Puccioni2015,Vallet2019}. Following the simple scheme of~\cite{Puccioni2015}, where all events (pump, spontaneous and stimulated emission and transmission through the cavity mirror) are treated as stochastic, integer variables based on a recurrence relation, we add the free propagation of the emitted photons and the reinjection of the chosen fraction (determined by $R_3$) through the output coupler ($M_2$) as another stochastic process. Thus, the external optical reinjection amounts to taking the sequence of emitted photons, and reinjecting the prescribed fraction on the basis of a probabilistic (Poisson) law~\cite{Wang2019a} with a delay which corresponds to the propagation time $2 \tau_{ext}$.

\section{Theoretical Model}
The simulated recurrence relations are those defined in~\cite{Puccioni2015} with the optical feedback added as in~\cite{Wang2019a}. Since only the on-axis stimulated photons and spontaneous photons can reenter the laser cavity, we just put the feedback on these two processes:
\begin{eqnarray}
\label{defN}
N_{q+1} & = & N_q + N_P - N_d - E_S \, , \\
S_{q+1} & = & S_q + E_S - L_S + S_{sp} + S_{inj,q-d}\, , \\
R_{L,q+1} & = & R_{L,q} + D_L - L_L - S_{sp} + R_{inj,q-d} \, , \\
\label{defRo}
R_{o,q+1} & = & R_{o,q} +(N_d - D_L) - L_o \, ,
\end{eqnarray}
where $N_P$ is the pumping process, $N_d$ represents the process of spontaneous relaxation which reduce the population inversion $N$, $E_S$ represents the processes of stimulated emission which also consume $N$, $L_S$ represents the leakage of stimulated photons through the output coupler, $S_{sp}$ is the seed starting the first stimulated emission process~\cite{Puccioni2015}, $D_L \propto \beta$ is the fraction of spontaneous relaxation processes which enter the on-axis mode (and therefore superpose to the stimulated emission), $L_L$ represents the losses for the on-axis fraction of the spontaneous photons through the output coupler, and $L_o$ the losses for the off-axis fraction of the spontaneous photons exiting (laterally) the cavity volume.  $S_{inj, q-d}$ and $R_{inj, q-d}$ represent the fraction of stimulated and spontaneous photons corresponding to the delayed index $(q-d)$ where $d$ is the index which matches the delay time $2 \tau_{ext}$. Detail parameters used in the simulation can be found in ~\cite{Puccioni2015}.
We define $\beta$ as the fraction of spontaneous emission coupled into the on-axis mode~\cite{Bjork1991}. The random processes are all defined as Poissonian distributions (which can also be implemented as Binomials~\cite{Puccioni2015} when additivity is fulfilled) with a probability which depends on the rate at which the described phenomenon occurs. 
The values of the rates used in the simulations are based on estimates for small nanopillar devices and amount to (cf.~\cite{Puccioni2015} for their role in the implementation in the stochastic scheme):   $\gamma_{\parallel} = 3 \times 10^9 s^{-1}$ for the spontaneous emission (process $N_d$), $\Gamma_c = 1 \times 10^{11} s^{-1}$ for the losses of the on-axis photons (i.e., stimulated, but also fraction of spontaneous photons in the lasing mode -- $L_S$ and $L_L$ processes), $\Gamma_o = 5 \times 10^{13} s^{-1}$ for the lifetime of off-axis (spontaneous) photons ($L_o$ process).  The pump rate $N_P$ represents the number of pumping processes in a cycle and, to maintain additivity for the statistical distributions, it is kept (for most draws) to either $0$ or $1$ by choosing a suitable time step.  $E_S$ is proportional to the product $\gamma_{\parallel} \beta S N$, as the standard probability of obtaining a stimulated process.  For consistency with the experiment conducted on a mesoscale laser~\cite{Wang2019a}, we keep $L_{ext} = 35 cm$.  The fraction $f$ of photons reinjected into the laser, and $\beta$ -- fraction of spontaneous emission coupled into the lasing mode --, are used as parameters in the simulations.
$S_{sp}$ is a flag used to initiate the stimulated emission (cf.~\cite{Puccioni2015} for details).  $S_{inj,q-d}$ and $R_{inj,q-d}$ is the fraction of back propagating photons which have probability $3 \times 10^{-3}$ of being reinjected into the cavity. The effective reflectivity $R_3$ determines the feedback fraction defined below.

Since at the present time only photon counters are sufficiently sensitive to measure the very weak output flux of nanolasers, one of the best indicators used to analyze the device's behaviour in experiments is the time-shifted second order correlation function~\cite{Wiersig2009} defined for the instantaneous photon number $M(t) = S(t) + R_L(t)$
\begin{equation}
\label{corrdef}
g^{(2)}(\tau) = \frac{\langle M(t)M(t+\tau)\rangle}{{\langle M(t) \rangle}^2}
\end{equation}
\noindent where $\langle M(t) \rangle$ denotes time-averaged photon number and $\tau$ the (variable) time-shift operation. The full functional dependence of $g^{(2)}(\tau)$ provides insight into the buildup of coherence due to the onset of stimulated emission after the lasing threshold has been crossed~\cite{Moody2018}. ``Coherence'' is attained when $g^{(2)}(\tau)$ approaches the Poisson limit ($g^{(2)} = 1$). In this case, the variance in the photon number is equal to that of a coherent state with the same mean photon number. On the other hand, the ``ideal thermal source'' is characterized by $g^{(2)}(\tau) = 2$, but we will consider $g^{(2)}(\tau) > 1$ to denote a photon statistics where ``thermal'' features still persist, i.e., where the variance in photon number is larger than for a coherent state~\cite{Moody2018}.  In the following, $g^{(2)}(\tau)$ will be the main indicator used for the analysis of the dynamics, even though we will also make use of others to help in the interpretation of its behaviour.  One of the scopes of this work is, indeed, a characterization of the potential for identifying the dynamical features with the help of correlations alone to help as guidance in future experiments.

\section{Results and Discussion}
We focus our numerical investigation on a $\beta = 10^{-1}$ nanolaser and introduce the feedback fraction parameter defined by 
\begin{equation}
f_{ext} = \frac{n_{in}}{n_{out}} \, ,
\end{equation}
where $n_{in}$ represents the number of photons coupled back into the laser while $n_{out}$ stands for the number of photons outcoupled from the cavity.  We explicitly introduce this parameter to mimic experimental setups where the reflectivity $R_3$ of mirror $M_3$ is fixed, while feedback control is achieved through an additional element (not included in the schematics of Fig.~\ref{Archtec}). All pump values are referred to the so-called {\it threshold pump} $P_{th}$, uniquely defined in~\cite{Rice1994} as $P_{th} = \frac{\Gamma_c}{\beta}$, which corresponds to the mid-point in the steep portion of the steady state curve representing photon number vs. pump.  As such, $P_{th}$ is a convenient, well-defined point to which all pump values can be normalized, even though in the context of a nanolaser, and at variance with macroscopic device, it does not correspond to a sharp threshold value.

\begin{figure}[ht]
\includegraphics[width=0.8\linewidth,clip=true]{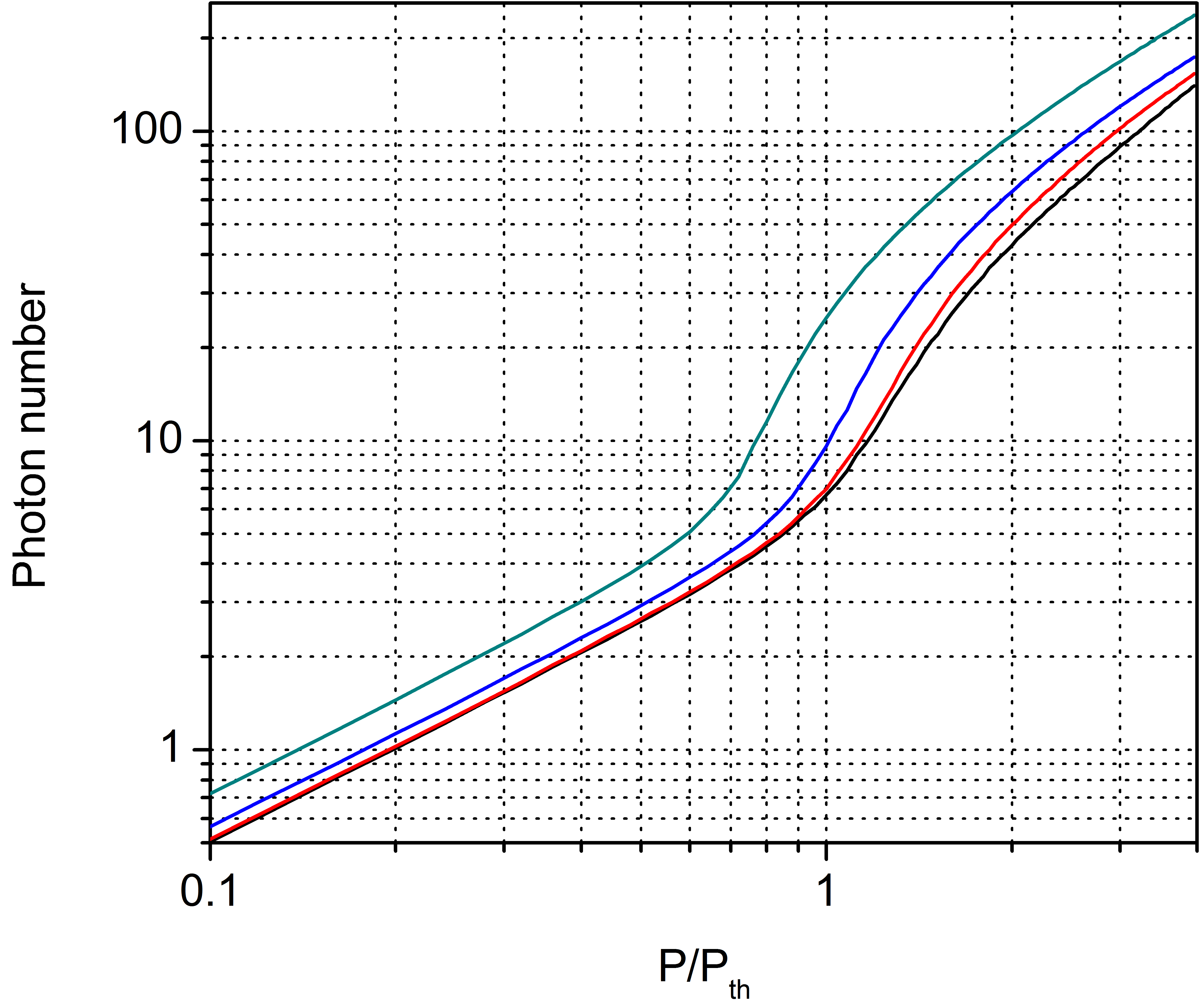}
\caption{Input-output laser response for different feedback fractions $f_{ext} =  0$ (free-running laser, black line); $0.015$ (red), $0.1$ (blue), $0.3$ (cyan).}
\label{IV-curve-log}
\end{figure}

Fig.\ref{IV-curve-log} shows the input-output response, for different values of $f_{ext}$, obtained by plotting the average output signal computed over $10\mu s$. For the solitary laser (black) the response shows the usual smooth growth of the photon number, with a broad transition region between the upper and lower branches.  At $f_{ext} = 0.015$ (red) no change is visible on the lower branch ($P/P_{th}<1$), but a deviation starts to appear for $P/P_{th}>1$, thanks to the minor, but non-negligible intensity contribution coming from the photons fed back into the cavity.  Further increasing the feedback fraction first enhances the effect above threshold (blue line), then displaces the whole response towards lower pump values (cyan line), thus effectively reducing the lasing threshold~\cite{Sondermann2003}.  In addition to the threshold shift, we also notice a larger differential increase (by approximately 30\%) in the above-threshold photon number, compared to its corresponding growth below threshold when comparing the cyan and black curves.  Thus, there is an overall deformation of the response curve, as if the laser's effective $\beta$-factor had been somewhat decreased.

\subsection{Characterization of the dynamics by the zero-delay autocorrelation}
\begin{figure}[ht]
\includegraphics[width=0.8\linewidth,clip=true]{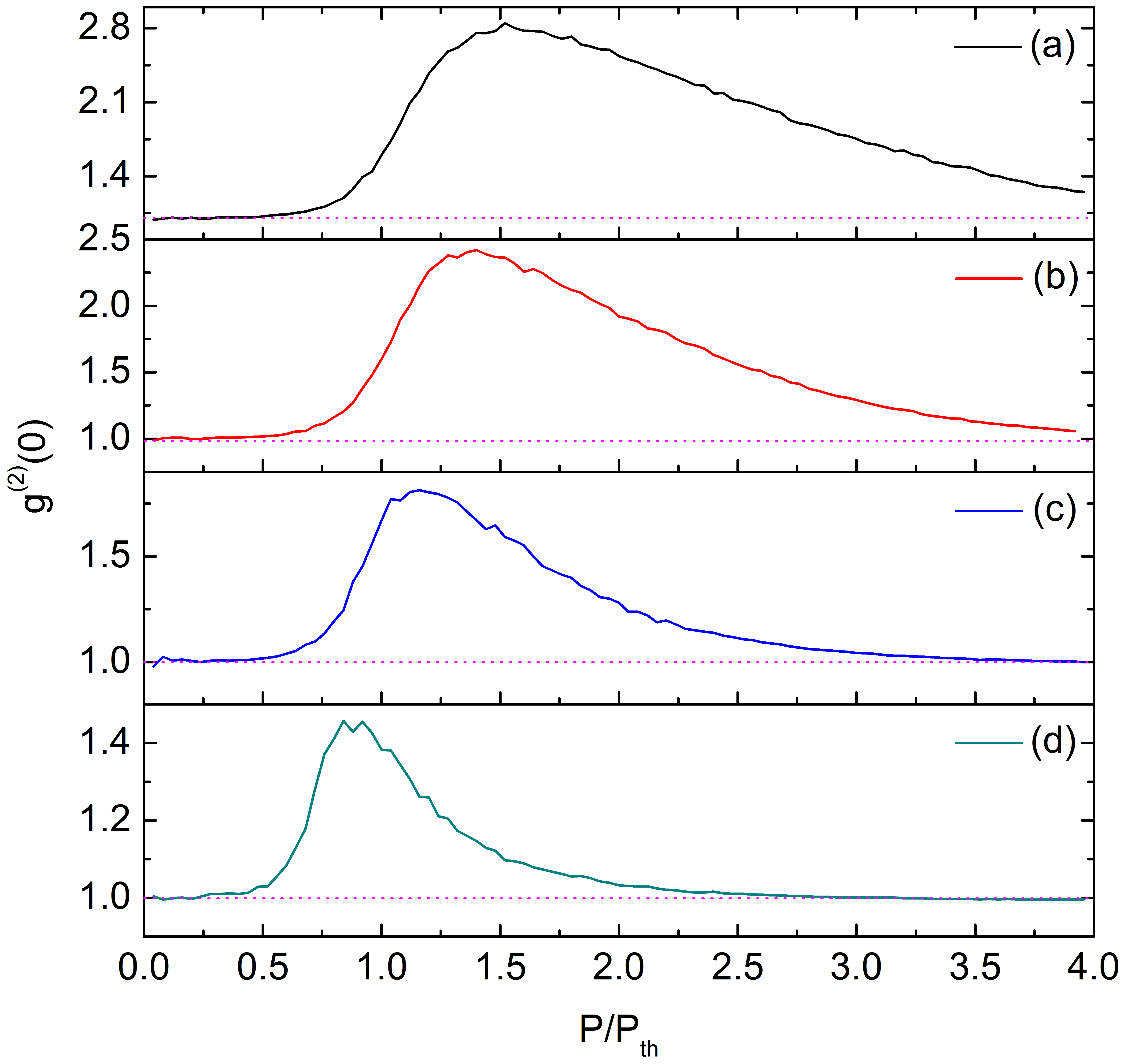}
\caption{Second-order autocorrelation function ($g^{(2)}(0)$) as function of normalized pump for $f_{ext} = 0$ (black curve); $0.015$ (red curve); $0.1$ (blue curve); $0.3$ (cyan curve).  The dotted horizontal lines (all panels) mark the $g^{(2)}(0) = 1$ level.}
\label{autocorrlation}
\end{figure}

Fig. \ref{autocorrlation} shows the zero-delay second order autocorrelation function curves ($g^{(2)}(0)$) for the different feedback levels. The value for the second-order autocorrelation below threshold falls towards 1, rather than the expected value 2 in the spontaneous emission regime, due to numerical difficulties.  These are encountered any time the direct estimate of $g^{(2)}(0)$ is computed from a (simulated) sequence of spontaneous emission events, more or less independently of the scheme used.  They stem from the amount of computing time that is required to collect sufficient statistics in a regime where very few events are recorded, due to the low excitation.  While it is in principle possible to obtain the thermal emission statistics, it becomes more and more numerically costly as the pump value is reduced, with computing times which can become of the order of many days or weeks. Thus, since there is no particular information to be gained from this effort, we keep the computing costs reasonable and accept the ``filtering'' action introduced by the insufficient sampling whose effect is, eventually, to produce shot-noise-like statistics ($g^{(2)}(0) = 1$). 

When the laser is operated with no feedback (black curve), $g^{(2)}(0)$  displays a slow decay starting from its maximum value ($2.8$), with super-Poissonian photon statistics ($g^{(2)}(0) > 1$) persisting for pump values larger than four times threshold.  Superthermal statistics $g^{(2)}(0) > 2$ appears at the peak~\cite{Jahnke2016} indicating photon bunching of dynamical nature, similar to what experimentally observed in a single-mode microcavity laser~\cite{Wang2017}.  The super-Poissonian statistics signals a late onset of ``coherence'' (defined as the cw emission of photons), albeit in highly variable numbers~\cite{Wang2018}), in spite of $g^{(2)}(0) > 1$.   

The addition of a small amount of feedback ($f_{ext} = 0.015$) (red curve), slightly accelerates the decay towards the limit $g^{(2)}(0) = 1$, which, however, is not reached until $P > 4 P_{th}$.  The autocorrelation peak -- somewhat smaller but still superthermal -- is attained at a lower pump value, but still well beyond $P_{th}$.  The decay in $g^{(2)}(0)$ is further accelerated by increasing the feedback fraction ($f_{ext} = 0.1$) and the Poisson limit is attained for $P \approx 3.5 P_{th}$, while the peak is now subthermal and occurs closer to $P_{th}$.  Finally, for $f_{ext} = 0.3$ we observe a subthermal autocorrelation peak at approximately $P_{th}$ with convergence to the Poisson statistics at $P/P_{th} \approx 2.5$.  The progressive shift in the peak position is consistent with the translation of the input-output nanolaser response (Fig.~\ref{IV-curve-log}) towards smaller pump values (threshold reduction).  As already discussed, the growth of the autocorrelation from the shot noise value for $P < P_{th}$ stems from the difficulty in collecting a sufficiently large number of events in a reasonable computing time below threshold.  

It is also important to remark that when the Poisson regime is attained, the model is expected to break down.  The strength of our current approach lies in its ability to capture the dynamics of intrinsic fluctuations, but, being based on a photon number description~\cite{Puccioni2015}, it cannot account for phase coherence.  As long as the field's coherence is smaller than the dephasing introduced by the feedback arm~\cite{Wang2019a}, the predictions will hold, but this will no longer be true once proper, noise-free lasing oscillation is in place.  At this stage, it is plausible to expect that in that case the phenomenology observed in macroscopic devices under the influence of feedback may hold, but this is a point that is left for further investigations. We currently focus on a pump range $P \leq 4 \times P_{th}$ chosen for the absence of coherence without feedback (Fig. 3a).  One must keep in mind that for the upper values of this pump range, for sufficiently strong feedback, the predictions may break down (reduction in signal complexity, cf. following figures). 

It is important, however, to realize that in the absence of feedback the emission consists mainly of spikes, similar to those observed in a lower $\beta$ device~\cite{Wang2015}.  In fact, the larger fraction of spontaneous emission coupled into the lasing mode extends the pump interval in which spiking is observed until a transition into noisy continuous oscillation (cw) is observed. For $\beta = 0.1$ we have estimated cw emission to appear for $P \approx 5 \times P_{th}$~\cite{Wang2018}, thus the reinjection consists of pulses which are uncorrelated from one another, lending more credibility to an entirely incoherent feedback scheme.

\subsection{Additional insight through temporal traces and power spectra}
\begin{figure}[ht]
\includegraphics[width=0.8\linewidth,clip=true]{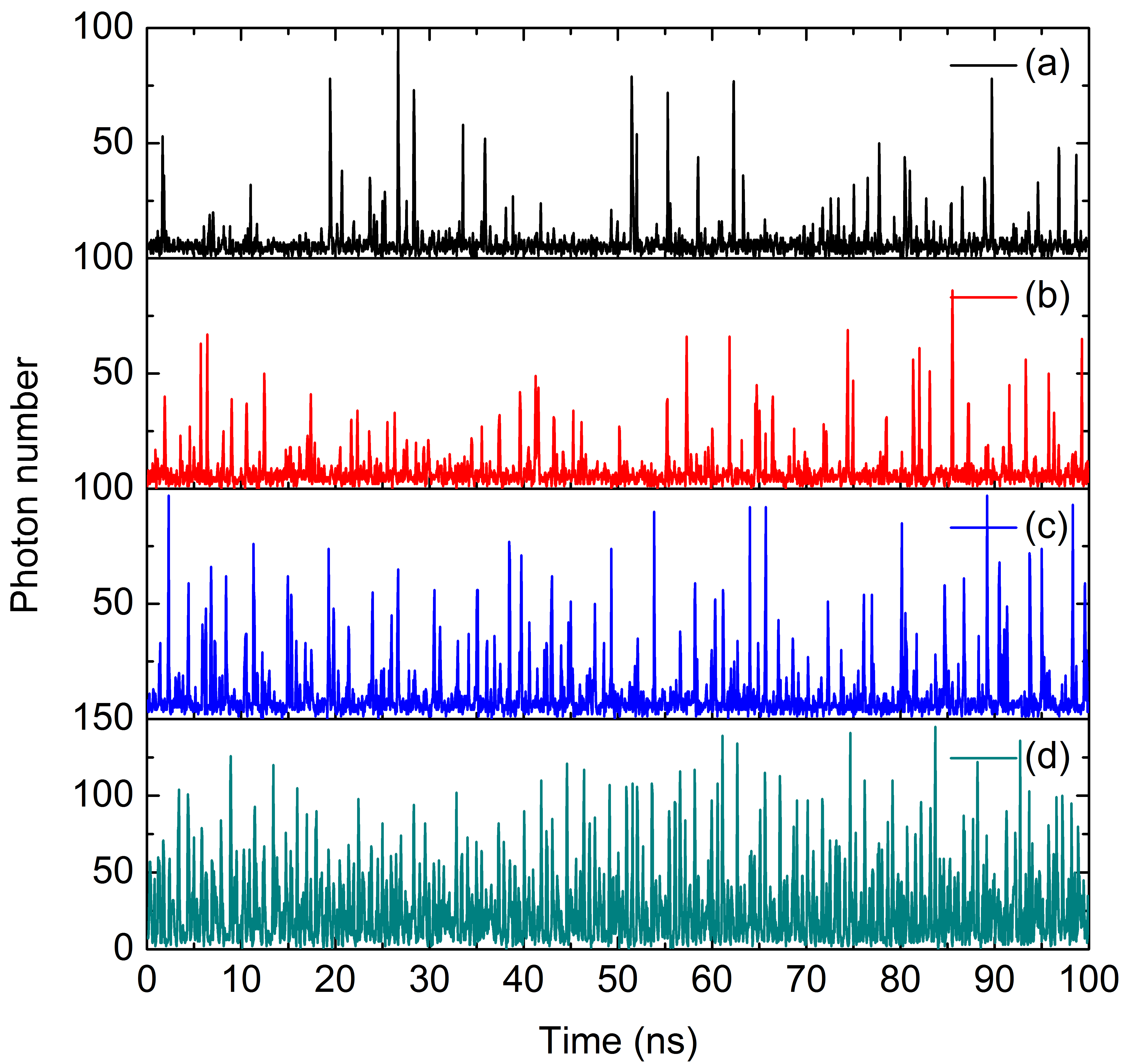}
\caption{Typical nanolaser dynamics with $f_{ext} = 0$(black curve); $0.015$ (red curve); $0.1$ (blue curve); $0.3$ (cyan curve).  $P = P_{th}$.}
\label{dynamics}
\end{figure}

The conclusion that one draws from the zero-delay autocorrelation is that feedback enhances the coherence, as signalled by the reduction in its peak height and in the earlier convergence towards the Poisson limit.  Unlike nanolaser experiments, the numerical simulation allows us to compare the previous results to the actual time traces, from which the autocorrelation is computed.  Fig.~\ref{dynamics} shows one representative sample of the temporal dynamics, displayed over a $100 ns$ time interval, observed outside the cavity (e.g. through the target, Fig.~\ref{Archtec}) when pumping at the (nominal) threshold point ($P = P_{th}$).

Overall, we observe a highly irregular spiking laser output both in the free-running case (Fig.~\ref{dynamics}a) -- known from previous simulations and from experiments in microcavity devices~\cite{Wang2015} -- as well as in the presence of feedback.  The main influence of feedback is to increase the number of spikes, without affecting their regularity, at least for low values of $f_{ext}$ (Fig.~\ref{dynamics}b). For $f_{ext} = 0.015$ the average height of the spikes is visibly reduced, thus explaining the very similar average between the two lowest values of $f_{ext}$ observed in Fig.~\ref{IV-curve-log} (compare black and red curves at $P = P_{th}$).  In other words, weak feedback appears to {\it regularize} somewhat the laser output by {\it redistributing} the emitted energy into smaller, more frequent photon bursts.  This observation qualitatively matches experimental measurements carried out in a microcavity device under comparably weak feedback (cf.~\cite{Wang2019a}, Fig. 4). 

At larger feedback fraction $f_{ext} = 0.1$ (Fig.~\ref{dynamics}c), we observe that irregular spikes have become mostly larger, while their average number (per unit time) appears similar to the one at weak feedback.  This suggests that the larger portion of reinjected photons strengthens the amplification process which has been advanced by the weak feedback.  This corresponds to the 40\% gain in average photon number observed in Fig.~\ref{IV-curve-log} when passing from weak to intermediate feedback (red and blue curves, respectively).  Finally, at $f_{ext} = 0.3$ we observe a strong enhancement in the frequency of the spikes accompanied by the onset of a non-sustained continuous component, which can be recognized by observing that the photon number trace does not touch the zero axis.  The establishment of lasing is well under way, as proven by the additional gain factor (approximately 3) relative to the intermediate feedback level (cf. Fig.~\ref{IV-curve-log} blue and cyan curves). 

\begin{figure}[ht]
\includegraphics[width=0.8\linewidth,clip=true]{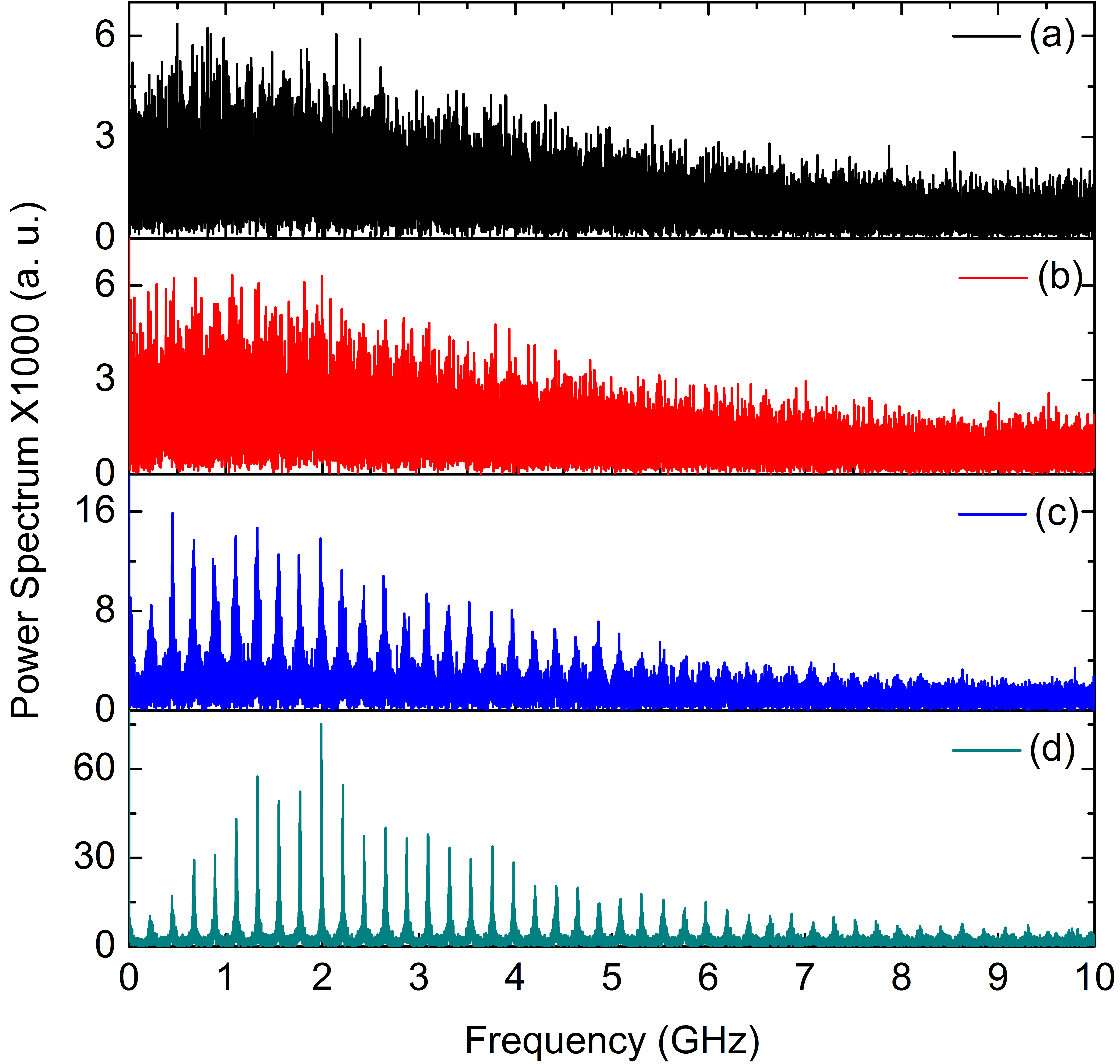}
\caption{Radiofrequency spectra computed on $\approx 4 \times 10^4$ points for $f_{ext} = 0$ (black curve); $0.015$ (red curve); $0.1$ (blue curve); $0.3$ (cyan curve).  $P = P_{th}$.}
\label{powerspectrum1}
\end{figure}

Complementary information can be gained from the corresponding radiofrequency (rf) spectra, computed at $P = P_{th}$ by taking the Fourier Transform of the temporal signal.  The spectra in the absence of feedback and for weak feedback look, at first sight, quite similar (Fig.~\ref{powerspectrum1}a,b).  Upon closer inspection one notices a more homogeneous distribution in the spectral densities for weak feedback which reflects the higher average spike frequency observed in the time traces.  

At intermediate feedback level (Fig. 5c) and thanks to the threshold lowering, the external cavity peaks appear in the time-delayed autocorrelation as satellites of the relaxation oscillation peak, with frequency differences matching the external cavity length~\cite{Huys2014}. It is important to remark that these peaks are well above the background, which maintains the level observed at weak (and no) feedback and extend quite far in frequency (at least to  $5 GHz$), thus showing the appearence of a periodicity driven by the external cavity which was not recognizable from the time traces.  Unlike the overall features remarked so far -- which can be shared by a free-running laser with lower threshold -- this is a specific contribution of feedback.  The peak heights, however, do no appear to possess strong overall structure in Fig.~\ref{powerspectrum1}c, save for a gentle decay at high frequencies and a very weak contribution at the fundamental (i.e., external cavity) component.

Interestingly, the rf spectrum acquires a structure which is much better defined for large feedback levels (Fig.~\ref{powerspectrum1}d).  While the peaks gain in strength, they also reflect the appearance of a resonance around the relaxation oscillation frequency ($\approx 2 GHz$) which is typical of above-threshold operation, even in a spiking regime~\cite{Wang2015}.  This matches the remark made when observing the temporal traces (Fig.~\ref{dynamics}d) that the output at times does not reach the zero photon number, i.e., a continuous output sets in (thus a low-coherence emission), in spite of its still strongly varying value.  Notice that the absolute value of $g^{(2)}(0) \approx 1.4$ is consistent with low-coherence but continuous output experimentally measured in a microcavity device~\cite{Wang2017}.  Finally, we remark that the harmonics also extend to even higher frequencies than for $f_{ext} = 0.1$.

Even though in this paper we are considering a nanolaser ($\beta = 0.1$), it is useful to draw some parallels with the experimental investigation carried out in the incoherent, low feedback regime in a microcavity device~\cite{Wang2019a}.  As already mentioned, the reduction in spike amplitude together with an increase in their average repetition rate qualitatively match the experimental observations (Fig. 4a in~\cite{Wang2019a}).  From the spectral point of view, we remark the absence of any structure in the spectrum (Fig.~\ref{powerspectrum1}b) as opposed to the emergence of a few peaks from the background reported in the microdevice (\cite{Wang2019}, Fig. 3a).  The discrepancy is likely to be related to the steeper response of a low-$\beta$ ($\approx 10^{-4}$) laser, as compared to the current $\beta = 0.1$ (cf.~\cite{Wang2016}, Fig. 2), which may induce effects more similar to those obtained at larger feedback here (e.g., compare Fig. 3a in~\cite{Wang2019a} to Fig.~\ref{powerspectrum1}c).  It is also interesting to remark that the resonance with the relaxation oscillations observed at large feedback for $P = P_{th}$ (Fig.~\ref{powerspectrum1}d) also appears in the microcavity experiment at larger pump for low feedback (Fig. 3c in~\cite{Wang2019a}), in agreement with the interpretation that $f_{ext} = 0.3$ plays the role of reducing the threshold, thus increasing the {\it effective} pump value at $P = P_{th}$.  This suggests a certain amount of genericity in the incoherent feedback scenario which goes beyond the specifics of the laser size. The larger range of pump values over which the transition between lower and upper branch occurs at large $\beta$ introduces a displacement in the parameter values where similar behaviour is observed and allows for the appearence of {\it intermediate} phenomena (such as the transformation of rarer and larger spikes into more frequent and smaller ones observed in the time traces, Fig.~\ref{dynamics}a,b) which probably go unnoticed in the larger devices. 

Before concluding this section, it is useful to recall the Wiener-Kinchin theorem:  the Fourier transform of the electric field autocorrelation $g^{(1)}(0)$ -- experimentally accessible with the help of an interferometric coincidence setup in a nanolaser experiment -- provides the power spectrum (or spectral density).  Thus the  information we have obtained from the rf spectra is accessible for an actual nanodevice.

The deeper insight provided by the analysis of the temporal traces (and power spectra) confirm the validity of the remarks made from the zero-order autocorrelation functions, which, however, give a much more limited amount of information.

\subsection{Characterization through the time-delayed autocorrelation}

\begin{figure}[ht]
\includegraphics[width=0.8\linewidth,clip=true]{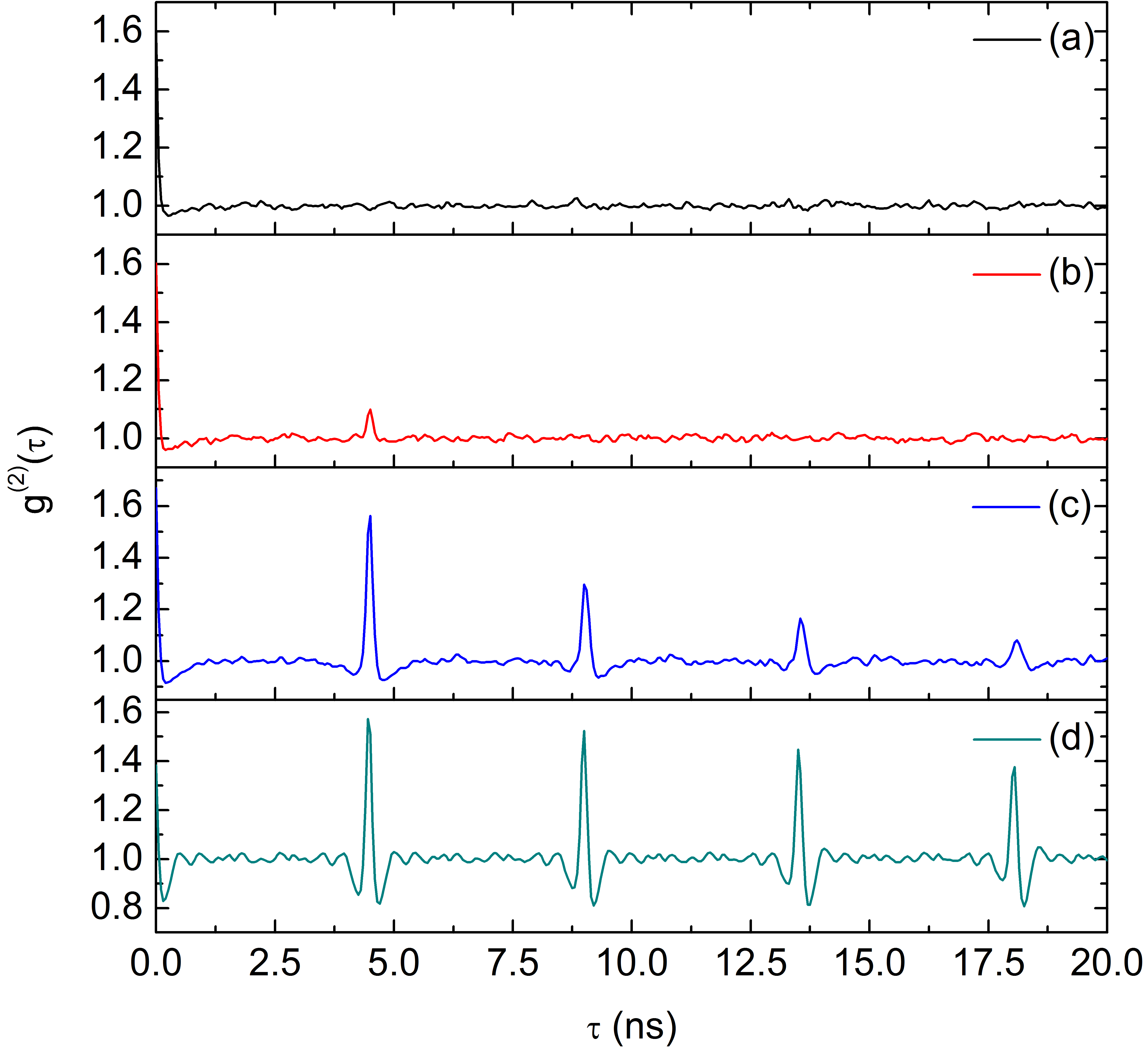}
\caption{Second-order correlation function $g^{(2)}(\tau)$ as function of the normalized pump for the laser operated freely (black curve), with $1.5\%$ feedback (red curve), with $10\%$ feedback (blue curve), and with $30\%$ feedback (cyan curve), respectively.  $P = P_{th}$.}
\label{correlationtime}
\end{figure}

An additional indicator currently available in experiments on nanolasers is the time-delayed second-order autocorrelation, which in our simulations can be computed directly from the temporal laser output (eq.~(\ref{corrdef})). Fig.~\ref{correlationtime} plots the autocorrelation for the different feedback levels previously analyzed (for the same pump value:  $P = P_{th}$).   Not surprisingly, in the free-running regime, no correlation is found (Fig.~\ref{correlationtime}a), while a very small peak appears at the roundtrip time of the external cavity for weak feedback ($f_{ext} = 0.015$, Fig. ~\ref{correlationtime}b).  This confirms the visual indication, Fig.~\ref{dynamics}b, that the spike emission is (almost) devoid of any regularity and follows for the most part the same random process observed in the free-running operation (panel (a)).  It also implies that almost all the rf spectral differences between $f_{ext} = 0.015$ and  $f_{ext} = 0$ (compare panels (a) and (b) in Fig.~\ref{powerspectrum1}) are only a reflection of the different spike amplitudes rather than actual spectral components -- up to the lowest portion of the spectrum $\nu < 0.25 GHz$.  

For moderate feedback levels ($f_{ext} = 0.1$, Fig.~\ref{correlationtime}c) peaks appear at multiple roundtrip times indicating the establishment of a degree of repetition, difficult to recognize in Fig.~\ref{dynamics}c, but visible in the rf power spectrum (Fig.~\ref{powerspectrum1}c) in the form of equispaced spectral peaks.  The amount of correlation decreases, however, fairly rapidly as a function of roundtrip number. Finally, at large feedback, $f_{ext} = 0.3$, sharp revivals at the cavity roundtrip time appear and extend well beyond the four periodicities shown in the graph (Fig.~\ref{powerspectrum1}d).  It is interesting to notice that now a structure develops around each revival peak with width $\Delta \tau \approx 2 ns$ containing a noticeable dip (below $g = 1$) and two clearly recognizable oscillations.  The resulting spacing ($0.5 ns$) matches the maximum height in the spectral peaks, Fig.~\ref{powerspectrum1}d, and indicates the presence of relaxation oscillations (at $\nu_{ro} \approx 2 GHz$).  Comparing to panel (c), which only showed a shallow and poorly defined dip, this change in functional structure signals the onset of coherence in the laser emission when passing from $f_{ext} = 0.1$ to $f_{ext} = 0.3$ in the form of a cw, even though very noisy, laser output.  Peaks are no longer disconnected from each other but form a {\it continuous} stream of photons whose numbers still strongly vary in time.  This observation rejoins the remark made when commenting the appearance of the temporal trace (Fig.~\ref{dynamics}d) which showed that the photon number often did not reach the $M = 0$ photons level.  However, the remark for Fig.~\ref{dynamics}d was purely qualitative and based on visual inspection (with limited resolution), while $g^{(2)}(\tau)$ provides a quantitative proof.  Interestingly, this also implies that $g^{(2)}(\tau)$ can provide equivalent information to that contained in the rf power spectrum, which is good news for experimentalists since the setup for the measurement of $g^{(2)}(\tau)$ is much simpler than the interferometric one needed for the measurement of $g^{(1)}$.

Comparing again to the experiment conducted in a microcavity device~\cite{Wang2019a}, we remark that the appearance of a structure around the revival peak at the roundtrip time in $g^{(2)}(\tau)$ had already been observed both in the experiment and in the numerical simulations (Figs. 6 and 9, respectively, in~\cite{Wang2019a}).  The not unexpected similarity confirms the usefulness of $g^{(2)}(\tau)$ as a tool for investigating the dynamics even at the nanoscale.

\begin{figure}[ht]
\includegraphics[width=1\linewidth,clip=true]{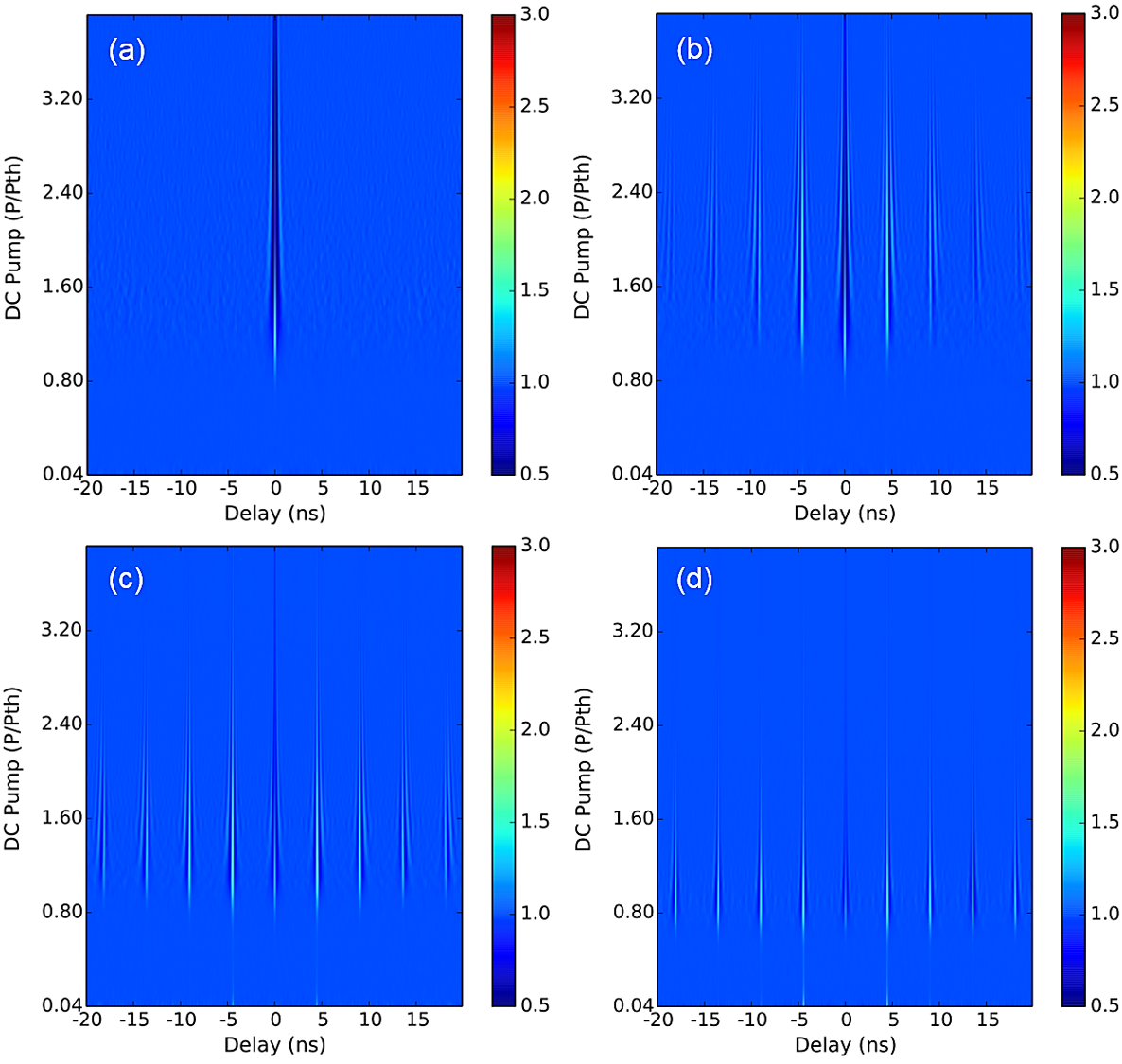}
\caption{Delayed second-order autocorrelation ($g^{(2)}(\tau)$, color scale)  for $\beta = 0.1$ and different pump values (vertical axes).  $f_{ext}$: (a) $0$; (b) $0.015$, (c) $0.1$, (d) $0.3$.}
\label{pump-correlation1}
\end{figure}

Complementary information can be gained from an overview of the time-delayed second-order autocorrelation over a range of pump values.  Fig.~\ref{pump-correlation1} plots the values of $g^{(2)}(\tau)$ in color scale whereby the vertical axis represents the pump and the horizontal the usual time delay $\tau$.  In the free-running regime, panel (a), we find another representation of Fig.~\ref{autocorrlation}a.  For weak feedback, instead, we observe the single revival which appears in Fig.~\ref{correlationtime}b at low pump, together with a second and a weak third as the pump grows.  Structuring around the first peak, in particular, is recognizable starting from $P \approx 1.6 P_{th}$ indicating again the appearance of a noisy, cw oscillation, in the same way that it was observed in Fig.~\ref{correlationtime}d. The same structure is clearly visible at moderate and high feedback (panels (c) and (d), respectively) but starting very close to threshold (or even for $P < P_{th}$ for $f_{ext} = 0.3$). However, the periodical dips have larger amplitudes when $f_{ext} = 0.3$, indicating a higher anticorrelation. We consider this may come from the more irregularity of the pulses. Finally, Fig. 7 shows the threshold reduction induced by feedback, but here the signal is not expected to be chaotic, at variance with ~\cite{Heiligenthal2013}, due to the strong stochastic component originating from the large portion of spontaneous emission coupled into the lasing more ($\beta = 0.1$). 

\begin{figure}[ht]
\includegraphics[width=0.8\linewidth,clip=true]{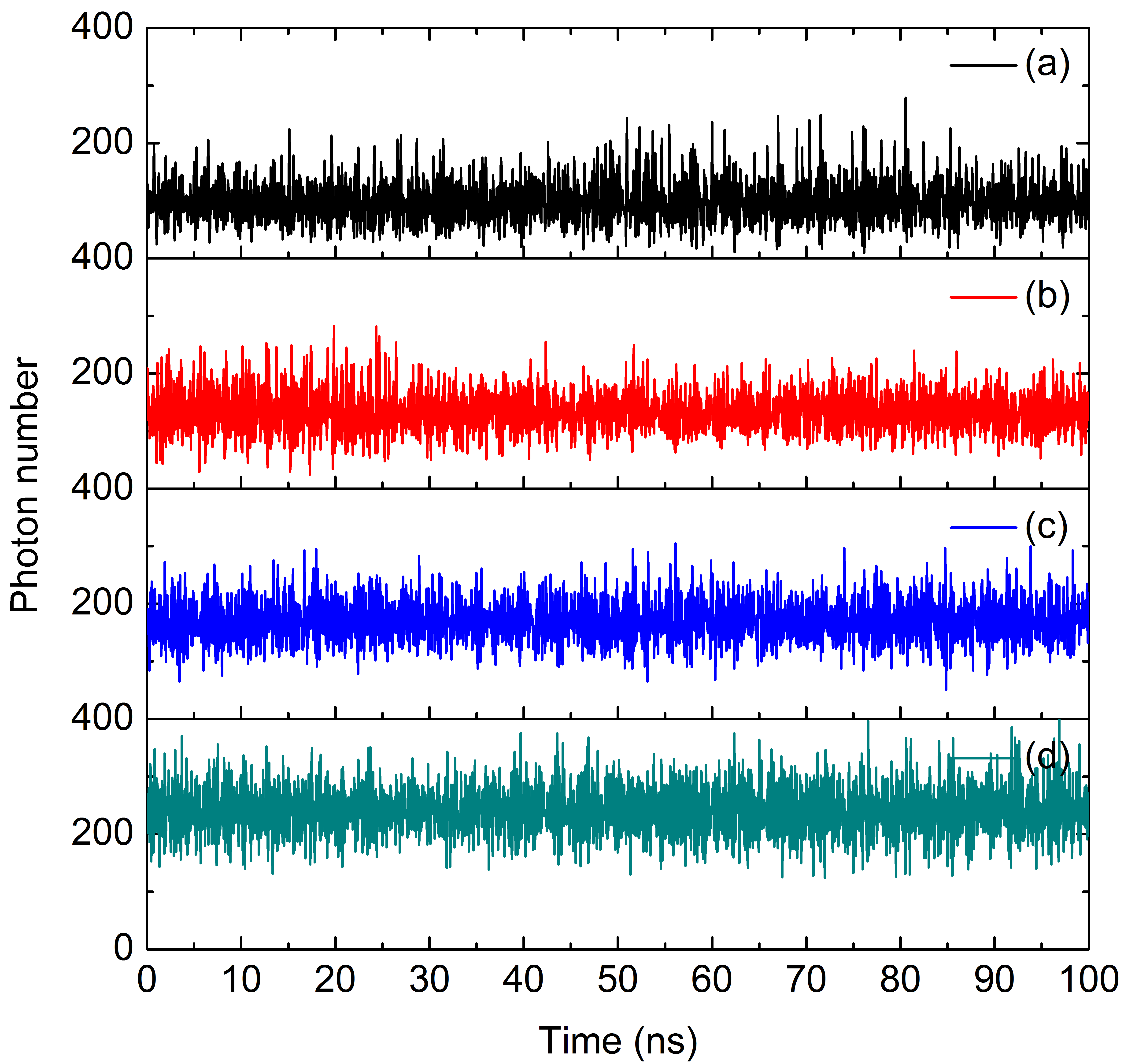}
\caption{Temporal dynamics of nanolaser with $10\%$ feedback $P = 2P_{th}$ (black curve),  $2.5P_{th}$ (red curve), $3P_{th}$ (blue curve), and $4P_{th}$ (cyan curve).}
\label{feedback-10percent}
\end{figure}

The last interesting feature of this overview is that it allows for an identification of the disappearance of spikes:  from $P \gtrapprox 1.6 P_{th}$ the correlation peaks nearly disappear at large feedback, signalling a strong reduction in the variability of the photon number, i.e., a small variance in the output, in agreement with $g^{(2)}(0) \lessapprox 1.1$ (Fig.~\ref{autocorrlation}d).  The same holds for the other feedback levels, for their corresponding pump values, as clearly illustrated in Fig.~\ref{feedback-10percent} for $f_{ext} = 0.1$:  while a spiking signal is still present at $P = 2 P_{th}$, already from $P = 2.5 P_{th}$ a cw oscillation is in place and evolves towards a noisy but certainly coherent emission at $P = 4 P_{th}$.

\section{Conclusion}

The results of this work extend a preceding investigation carried out on a $\beta \approx 10^{-4}$ microlaser subject to low-level optical feedback~\cite{Wang2018ieee}, comparing experiments and stochastic modeling.  The consideration of a nanolaser $\beta$ value is devoted to explore the predictions of the stochastic simulation, whose predictions have been shown to hold quite well for most of the investigated aspects of the dynamics~\cite{Wang2018ieee}, to nanodevices.  A good degree of qualitative similarity is observed between the two $\beta$ values, but for three main differences:  a steady-state response (S-curve) which is more differentiated as a function of feedback level, the appearence of a regime where irregular pulses become smaller and more frequent as the feedback grows, and the need for a substantially higher feedback level for qualitatively similar predictions.  While the general features remain the same, we also remark how predictions for the larger $\beta$ device possess more uniform features (e.g., in the spectral peak distribution in frequency space) than in the larger device (smaller $\beta$).
This recognition highlights the fact that the cavity volume plays an important role in the dynamics induced by incoherent feedback, characterized by photon spikes of shorter duration -- as visible both from time traces and power spectra -- and of lower amplitude, due to the overall smaller photon number at comparable pump values (relative to the respective laser thresholds).

From the predictions it is also clear how the nanolaser is much less sensitive to feedback-induced perturbations.  The need for higher feedback levels -- by at least one order of magnitude -- required to produce similar behaviour clearly indicates that a nanolaser should be capable of withstanding larger fractions of parasitic light reinjected into its cavity than a macroscopic, or even a mesoscopic~\cite{Wang2018ieee} laser.  From an application point of view this is very good news, since it reduces the need for optical isolation, a component which would be difficult, and expensive, to integrate into a chip.  It has been known for some time that Quantum Dots are much less sensitive to feedback, even at the macroscopic scale~\cite{OBrien2004,Huyet2004,Otto2010,Hopfmann2013} and certainly more so for nanodevices.  In addition, particular cavity configurations, such as that of a Fano laser, lend themselves to an exceptionally high degree of stability with respect to feedback~\cite{Rasmussen2019}.  However, our predictions are computed for a basic Quantum Well laser and show that even for this technology smaller devices are less sensitive to optical reinjection.

The second important aspect of the investigation concerns the use of the second order correlation function as a sole tool for the characterization of the dynamics.  While autocorrelations are known to be powerful tools, nonetheless the information they convey is based on the statistical collection of correlation events.  Thus, the question as to whether they would be able to distinguish the dynamical regimes holds true.  Through comparison between the autocorrelation predictions (without and with delay) with power spectra and temporal signals, we have shown that the principal features could be well captured, thereby validating the usefulness of the technique.  This is a reassuring point, since experimentally it is for the moment nearly impossible to gain direct information on nanolaser dynamics.

\section{Acknowledgments}
The authors are grateful to the R\'egion PACA and BBright for support. T. W. thanks the National Natural Science Foundation of China (61804036), the Open Foundation of the State Key Laboratory of Morden Optical Instrumentation at Zhejiang University, and Scientific Research starting fund (KYS045618036).  G.L. L. acknowledges discussions with T. Ackemann, M. Giudici, and S. Reitzenstein.  





\end{document}